\documentstyle[twocolumn,prl,aps,epsfig]{revtex}
\def\inseps#1#2{\def\epsfsize##1##2{#2##1} \centerline{\epsfbox{#1}}}
\begin{document}
\draft
\input{psfig}
\title{Two-scale competition in phase separation with shear}
\author{F. Corberi}
\address{Dipartimento di Scienze Fisiche, Universit\`a di Napoli and
Istituto Nazionale per la  Fisica della Materia, Unit\`a di Napoli,
Mostra d'Oltremare, Pad.19, 80125 Napoli, Italy}

\author{G. Gonnella and A. Lamura}
\address{Istituto Nazionale per la  Fisica della Materia,
{\rm and} Dipartimento di Fisica, Universit\`a di Bari, {\rm and}
Istituto Nazionale di Fisica Nucleare, Sezione di Bari, via Amendola
173, 70126 Bari, Italy.}
\date{\today}
\maketitle
\begin{abstract}
The behavior of a phase separating binary mixture 
in uniform shear flow is investigated by numerical simulations and
in a renormalization group (RG) approach.
Results show the simultaneous existence of domains of
two characteristic scales. Stretching  
and cooperative ruptures of the network produce a rich
interplay where the recurrent prevalence of thick and
thin domains determines log-time periodic oscillations.
A power law growth $ R(t) \sim t^{\alpha }$ of the
average domain size, with $\alpha =4/3$ and $\alpha = 1/3$
in the flow and shear direction respectively, is shown to be obeyed.

\end{abstract}


The application of a shear flow to a disordered binary mixture
quenched into a coexistence region greatly affects the phase-separation 
process \cite{On97}.
A large  anisotropy is  observed in typical patterns of domains
which appear greatly elongated in the direction of the flow
\cite{Has}.
This behavior has consequences for the rheological properties 
of the mixture.
A rapid strain-induced thickening followed by a gradual thinning
regime is observed \cite{KSH}.
Various numerical simulations confirm these observations 
\cite{OND,Ro,PT}.

In a recent paper \cite{PRL} we have studied the phase-separation
kinetics in the context of a self-consistent approximation,
also known as large-$N$ limit.
Within this approach the existence of a scaling regime
characterized by an anisotropic power-law
growth of the average size of domains was established.
The segregation process, however, cannot be fully described 
by this technique  
because interfaces are absent for large-$N$ \cite{B94}.

The aim of this letter is to investigate the ordering 
process by extensive numerical simulations. 
Our main result concerns the simultaneous existence
of {\it two} length scales which characterize the 
thickness of the growing domains. The competition between these scales 
produces a rich dynamical pattern with an  oscillatory behavior
due to the cyclical prevalence of one
of the two lengths.
A power law growth 
$R(t)\sim t^{\alpha }$ of the
average domain size, with $\alpha =4/3$ and $\alpha = 1/3$
in the flow and shear direction respectively,   
is shown to be obeyed by a renormalization group (RG)
analysis.

The kinetic behaviour of the binary mixture is described 
by the Langevin equation
\begin{equation}
\frac {\partial \varphi} {\partial t} + \vec \nabla (\varphi \vec v) =
\Gamma \nabla^2  \frac {\delta {\cal F}}{\delta \varphi} + \eta
\label{eqn2}
\end{equation}
where the scalar field $\varphi$ represents the concentration
difference between the two components of the mixture
\cite{On97}.  
The  equilibrium
free-energy can be chosen as usual to be 
\begin{equation}
{\cal F}\{\varphi\} = \int d^d x 
\{\frac{a}{2} \varphi^2 + \frac{b}{4} \varphi^4 
+ \frac{\kappa}{2} \mid \nabla \varphi \mid^2 \}
\label{eqn1}
\end{equation}
where  $b,\kappa >  0 $ and $a<0$ in the ordered phase. 
$\vec v$ is an external  velocity field describing plane shear flow
with average profile given by
\begin{equation}
\vec v = \gamma y \vec e_x
\label{eqn4}
\end{equation}
where $\gamma$ is the  shear rate 
and $\vec e_x$ is the  unit vector in the flow direction.
$\eta$ is a gaussian white noise,  representing thermal fluctuations,
with mean zero and correlation
$\langle \eta(\vec r, t) \eta(\vec r', t')\rangle = -2 T \Gamma \nabla^2 \delta(\vec r - 
\vec r') \delta(t-t')$, 
where $\Gamma$ is a mobility coefficient, $T$ is the 
temperature of the fluid, and the symbol $\langle ...\rangle $ denotes the 
ensemble average.

The Langevin equation (\ref{eqn2}) has been simulated in $d=2$
 by first-order Euler discretization scheme.
Periodic boundary conditions have been used in the flow direction
while in the $y$ direction the point at $(x,y)$ is
identified with the point at $(x+\gamma L, y+L)$, 
where $L$ is the size of the lattice \cite{OND}. 
Lattices with $L=1024, 2048, 4096$ and space discretization
intervals $\Delta x =0.5, 1$ were used. 
In the phase separation process  
the  initial configuration of $\varphi$ is  
a high temperature disordered state and
the evolution of the system is studied in  model  (\ref{eqn2}) with $a<0$.
Parameterization invariance of (\ref{eqn2}) allows one to set 
$\Gamma=|a|=b=\kappa=1$. 
The structure factor  is defined as 
$C(\vec k,t) = \langle \varphi(\vec k, t)\varphi(-\vec k, t)\rangle $ where 
$\varphi(\vec k, t)$ are the Fourier components of $\varphi$. 
Results will be shown for the case
$\gamma=0.0488$, $T=0$, $L=4096$, $\Delta x = 1$, 
$\langle \varphi  \rangle = 0 $.
Similar results have been obtained for other values of the parameters.

A sequence of configurations at different values of the strain $\gamma t$ 
is  shown in Fig.~1.
After an early time, when well defined domains are forming, the usual
bicontinuous structure of phase separating domains starts to be distorted
for $\gamma t \gtrsim 1$. 

\begin{figure}
\vskip 10mm
\inseps{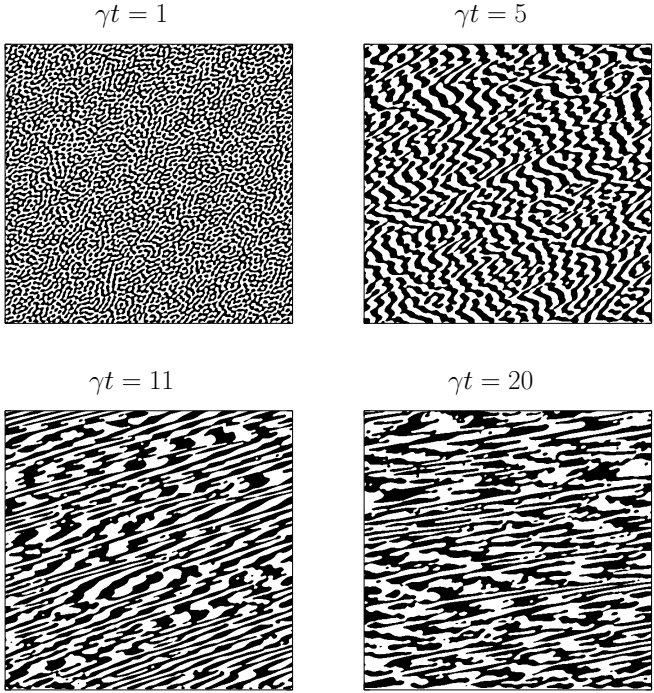}{0.6}
\vskip -70mm
\caption{Configurations of a portion of $512 \times 512$ 
sizes of the 
whole lattice are  shown at different values of the strain $\gamma t$.} 
\end{figure}

The growth is faster in the flow direction and domains assume the typical
striplike shape aligned at an angle $\theta (t)$ with  the direction of 
the shear which  decreases  with time.
As the elongation of the domains increases, nonuniformities 
appear in the system:
Regions with domains of  different thickness can be  clearly observed  
at $\gamma t =11$.  
The evolution at  still larger values of the strain
is shown for $\gamma t = 20$.
The  domains with the smallest thickness
eventually break up and burst with the formation of  
small bubbles.

A  systematic existence of two scales in the size distribution of domains
is suggested  by the behavior of the structure factor, shown in Fig.~2.
At the beginning (see the picture at $\gamma t =0.2 $)  $C(\vec k,t)$
exhibits an almost 
circular shape, corresponding to the 
early-time regime without sharp interfaces. 
Then  shear-induced anisotropy becomes evident,
$C(\vec k,t)$  is deformed into an ellipse, changing also its profile and,
for  $\gamma t\gtrsim  1$, 
four  peaks can be clearly  observed.
The position of each peak identifies a couple of typical lengths, one
in the flow and the other in the shear direction. The peaks are
related by the $\vec k \rightarrow -\vec k$ symmetry 
so that, for each direction, there are two physical lengths.
This  corresponds to the observation of domains with two 
characteristic thicknesses, made in  Fig.~1.
 
The dips in the profile of $C(\vec k,t)$ develop with time until 
$C(\vec k,t)$ results  to be separated in two distinct foils 
at  $\gamma t \simeq 4 $.
The evolution of the system until this stage is well described by the 
solution of the linear part of  (\ref{eqn2}) (letting $b=0$):
\begin{equation} 
C(\vec k,t) = C_0 \exp^{-\int_0^t  k^2(s)( k^2(s) - a) ds}
\label{eqnls}
\end{equation}
where $\vec k(s) = (k_x, k_y +\gamma s k_x)$ and $C_0$ is the structure factor
at the initial time \cite{Dino}. 
Then non-linear effects become essential in producing the patterns
shown    in Fig.2 at $\gamma t = 11, 20$.
At $\gamma t = 11 $ we evaluate the positions of the peaks at
$(k_x,k_y)=(0.015,0.107)$ and $(k_x,k_y)=(0.038,0.23)$.
This gives a value around two for the ratio between the characteristic
sizes of domains;  the 
 same value  is found  for the ratio between the positions of the two 
peaks in the hystogram of the domain size  distribution.
The relative height of the peaks in one of the foils
of $C(\vec k,t)$  can be more clearly
seen in Fig.3 where  
the two maxima  are observed to dominate alternatively at the 
times $\gamma t = 11$ and $\gamma t = 20$.

\begin{figure}
\epsfig{file=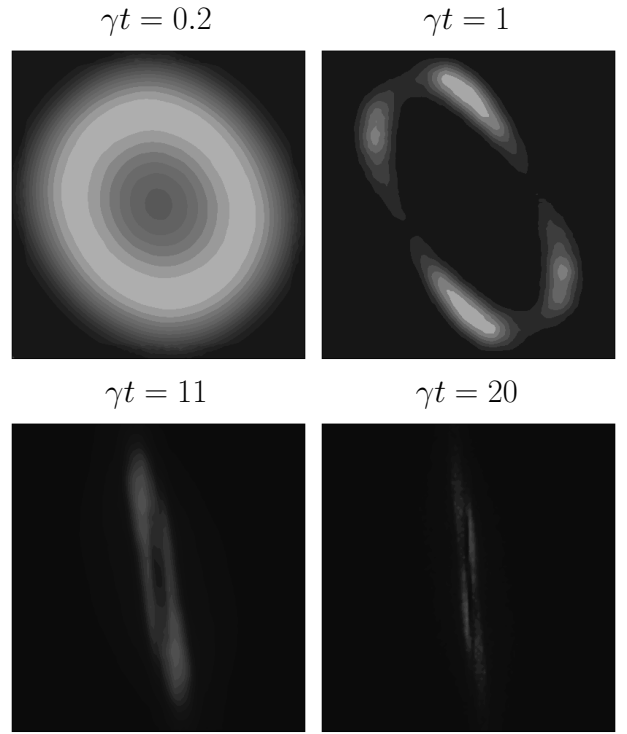,bbllx=105pt,bblly=425pt,bburx=430pt,bbury=825pt,width=8.2 cm,clip=}
\caption{The structure factor is shown at different values 
of the strain $\gamma t$.} 
\end{figure} 

The competition between two kind{\bf s} of domains is  
a cooperative phenomenon. In a situation   like that at $\gamma t =11 $,
the peak with the larger $k_y$ dominates, describing a prevalence 
of stretched thin domains. When the strain becomes larger, 
a cascade of ruptures occurs
in those regions of the network where the stress is higher
and elastic energy is released. 
At this point the thick domains, which have not yet been broken,
prevale and the other peak of  $C(\vec k,t)$ dominates, 
as at $\gamma t=20$.
 In the large-$N$ limit the prevalence of one or the other peak has 
been shown to continue periodically in time \cite{PRL}. 
Here we have an indication 
of a similar behavior,  although the 
   observation of the recurrent dominance of the peaks on longer timescales 
   is hardly accessible numerically.

\begin{figure}
\epsfig{file=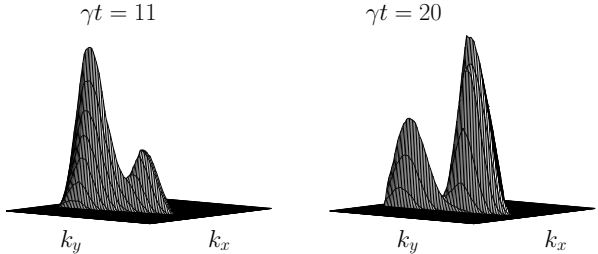,bbllx=70pt,bblly=530pt,bburx=505pt,bbury=730pt,width=8.2 cm,clip=}
\caption{The same structure factor of Fig.2 at $\gamma t =11, 20$
in a three-dimensional plot.}
\end{figure}

The hallmarks of this dynamics are found in the behavior
of the average size of domains, $ R_x(t)$ and $ R_y(t)$,
in the flow and shear direction.
In our simulations these quantities have been calculated
through
$  R_x(t) = ( \int d\vec k k_x^2 C(\vec k,t) / \int d\vec k  
C(\vec k,t))^{-1/2}$, and 
analogously for $ R_y(t)$; their behavior is shown in Fig.4.  
Due to the alternative dominance of the peaks of  $C(\vec k,t)$,
$ R_x$ and $ R_y$ increase with an oscillatory pattern.
The latter reaches
a local maximum at the characteristic times $t_n$,
when  $C(\vec k,t)$ is of the form of Fig.~2 at $\gamma t = 20$ and
thick domains are more abundant.
The time interval $\tau _n=t_{n+1}-t_n$ between two cascades of ruptures 
increases exponentially
as the phase separation proceeds so that the oscillations appear 
to be  periodic on a  logarithmic time scale. 
This is expected because domains are growing and longer and longer times 
are required to break them.
The following argument simply illustrates the origin of the log-time
periodic oscillations:
The break up of an elongated domain is caused by the shear flow,
which enters Eq.~(\ref{eqn2}) through the last term on the l.h.s.,
whose magnitude we infere to be  proportional to $\tau _n ^{-1}$.
For a sharp interface exposed with an angle $\theta (t)$ to the flow
this term is proportional to $\gamma \sin \theta (t)\simeq
\gamma \theta (t) \sim \gamma  R_y(t)/ R_x(t)$,
where  the asymptotic smallness of $\theta (t)$  has been taken into 
account.
The RG argument developed below shows that $\gamma R_y(t)/R_x(t)\sim 1/t$,   
so that $\tau _n \sim t_n$ and hence $\ln t_n\sim n+cost.$:
The rupture events occur periodically in $\ln t$.

Stretching of domains requires work against surface tension and results
in an increase  $\Delta \eta$ of the viscosity \cite{KSH,Onu}. 
We calculate  the excess viscosity as  
$\Delta \eta (t)= -\gamma^{-1} \int 
\frac {d\vec k}{(2\pi)^D} k_x k_y C(\vec k,t) $ \cite{On97}. 
Starting from zero  $\Delta \eta$ grows oscillating up to 
global maximum at $\gamma t \simeq 12 $.
Then the excess viscosity relaxes  to zero
in a similar way to what observed in previous simulations \cite{OND}.
The relative maxima of   $\Delta \eta$ are found in 
correspondence 
of the minima of  $ R_y$  when the domains are maximally 
stretched.
 
In the case without  shear the asymptotic kinetics is characterized by
dynamical scale invariance, which is reflected by a power law growth
$ R(t)  \sim t^{\alpha}$ of the average 
size  $ R(t)$ of domains \cite{B94}. 
The value of the growth exponent 
$\alpha$ is related to the physical mechanism operating 
in the separation process and is $\alpha=1/3$ for diffusive 
growth.
In the case with shear the existence of a similar behavior 
has not been clearly assessed. A stationary state
characterized by domains of finite thickness is generally  
observed in experiments \cite{mat}. Only in some
experimental 
realizations the existence of a
regime with power law growth has been shown \cite{Bey,LLG}.

The self-consistent solution of Eq.(\ref{eqn2}) shows \cite{PRL}
 that a generalized scaling symmetry holds when a shear flow is applied
and different exponents are found for the power-law growth of 
$ R_x(t)$ and $ R_y(t)$.
However these exponents  
are correct  for  models with  vectorial order parameter
and do not directly apply to the case of a binary mixture \cite{zan}.
The existence of a scaling symmetry and the actual
value of the growth exponents cannot be reliably studied by direct numerical
simulation of Eq.(\ref{eqn2}): This is because, even   
for the bigger lattice we have used, 
the fast growth in the $x$-direction makes finite size effects 
relevant before a full realization of the scaling regime occurs.
Moreover, the oscillatory behavior of $ R_x(t)$ and 
$R_y(t)$ prevents
a straightforward computation of the growth exponents, unless the evolution
of the system is followed over several decades, which is, presently, 
impossible.
Therefore, in order to infer the actual value of the growth exponents
we resort to a RG analysis.

As usual, we define the RG transformation for the Fourier components 
$\varphi_{\vec k}(t) =  1/\sqrt{V} \int
 {d\vec r} \varphi(\vec r,t) e^{i \vec k \cdot \vec r}$ 
of the field $\varphi(\vec r,t)$.
They verify the equation
\begin{equation}
\frac {\partial \varphi_{\vec k}(t)} {\partial t} - 
\gamma k_x \frac {\partial \varphi_{\vec k}(t)} {\partial k_y} =    
- \Gamma k^2 \frac {\delta  {\cal F}} {\delta \varphi_{-\vec k}(t)}
+ \eta_k(t)
\label{Lan}
\end{equation}
with
$ < \eta_k(t) \eta_{k'}(t') > = 2 T \Gamma k^2 \delta(t-t')
\delta(\vec k + \vec k')$.
Taking into account the anisotropic growth of domains, 
we  generalize the RG scheme of  \cite{RG}
by considering  the change of scale \cite{nota2d}
\begin{eqnarray}
k_x \rightarrow k'_x = k_x b^{\alpha_x}&,& \qquad
k_y \rightarrow k'_y = k_y b^{\alpha_y}, \nonumber \\
t &\rightarrow& t' = t b^{-1}
\label{eqn7}
\end{eqnarray}
and the field transformation
\begin{equation}
\varphi_{\vec k}(t) = b^{\zeta} \varphi'_{\vec k'}(t')
\label{eqn8}
\end{equation}
where $b$ is the rescaling factor and the meaning of the 
exponents $\alpha_x, \alpha_y, \zeta$ will be clarified in the following.
Dimensional analysis implies that the  structure factor  
can be written as 
\begin{equation}
C(\vec k,t) =  R_x(t)  R_y(t) f(x,y)
\label{eqc}
\end{equation}
with
$x=k_x   R_x(t), y= k_y  R_y(t)$. 
Its invariance in form with respect to the 
transformations (\ref{eqn7},\ref{eqn8}) gives 
$\zeta= (\alpha_x + \alpha_y)/2$. 
The invariance of the scaling variables $x,y$ under the
transformations (\ref{eqn7}) implies that $\alpha_x, \alpha_y$ 
correspond to  the growth exponents.
 
\begin{figure}
\epsfig{file=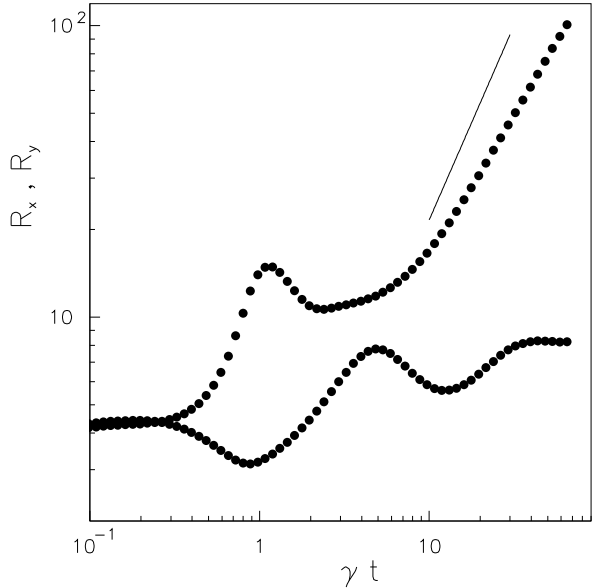,bbllx=15pt,bblly=130pt,bburx=530pt,bbury=655pt,width=8.2 cm,clip=}
\caption{Evolution of the average domain size
in the shear (lower curve) and flow  (upper curve) directions.
The straight line has slope 4/3.}
\end{figure}

The dynamical scaling regime is described by  a fixed point
of the Langevin  equation (\ref{Lan}) 
under the set of transformations (\ref{eqn7},\ref{eqn8}) 
\cite{Nota1}.
Generalizing the prescription  of \cite{RG}, we 
assume that the   fixed point hamiltonian ${\cal F}$
verifies the relation 
\begin{equation}
{\cal F}(b^{\zeta} \varphi'_{\vec k'}(t')) = b ^{\alpha_x } 
{\cal F}(\varphi'_{\vec k'}(t'))  
\label{eqn9}
\end{equation}
meaning that the main contribution to the free energy 
is due to the interfaces in the direction of the flow.
By inserting (\ref{eqn7},\ref{eqn8}) into (\ref{Lan}) 
and using (\ref{eqn9}), one obtains
a  Langevin  equation similar in form to (\ref{Lan}) with rescaled parameters
\begin{eqnarray}
\frac {1}{\Gamma'} = \frac{1}{\Gamma} b ^{2\zeta + 2\alpha_y -1 - \alpha_x }&,&
\qquad
\gamma' = \gamma b^{\alpha_y - \alpha_x + 1}, \\
T' &=& T b^{-\alpha_x }
\label{eqn10}
\end{eqnarray}
A  fixed point of the above recursions 
with $\Gamma, \gamma \ne 0$ is obtained when 
$\alpha_y = 1/3, \alpha_x = \alpha_y + 1$, with the temperature being
not  relevant  for the process of phase separation.  
We observe that a difference between the growth exponents
$\Delta \alpha = \alpha_x - \alpha_y$ in the range
$0.8\div 1$ has been measured in \cite{Bey,LLG}.
Moreover, the above analysis suggests that,
 for a constant value of the strain $\gamma t$,
the excess viscosity scales as   $\Delta \eta \sim \gamma ^{-\beta}$
with $\beta=1/3$.

In conclusion, we have studied the phase-separation kinetics of
a binary fluid in an uniform shear flow by direct numerical simulation
of the constitutive equations and in a RG approach.
Results show the simultaneous existence of domains of two characteristic sizes
in each direction. The two kind of domains alternatively prevail,
because the thicker are thinned by the strain and the thinner are
thickened after  cascades of ruptures in the network. This mechanism 
produces an oscillation which decorates the  expected power-law 
growth $ R(t)\sim t^\alpha$ of the average size of the domains, 
with $\alpha =4/3$ and $1/3$ in the flow and shear direction, respectively.
The oscillations occur on  logarithmic time-scale as in models
describing propagation of fractures in materials where the releasing of 
elastic energy is measured \cite{Sor}. Finally, in a recent paper \cite{July},
a first treatment of the effects of
 hydrodynamics 
  on the phase separation of a binary mixture in uniform shear
has been given. It would be interesting to 
 know if and how hydrodynamics 
 affects the global picture described in this letter.
 
~\\
We thank Dino Fortunato and Julia Yeomans for helpful discussions. 
F.C. is grateful to M.Cirillo and R. Del Sole 
for hospitality in the University of Rome.
F.C. acknowledges support by the TMR network contract ERBFMRXCT980183
and by MURST(PRIN 97).


\begin{references}

\bibitem{On97}
For a review, see A. Onuki, J. Phys.: Condens. Matter {\bf 9} 6119 (1997).

\bibitem{Has}
T. Hashimoto, K. Matsuzaka, E. Moses, and A. Onuki, Phys. Rev. Lett. {\bf 74} 
126 (1994).

\bibitem{KSH}
A.H. Krall, J.V. Sengers, and K. Hamano, Phys. Rev. Lett. {\bf 69} 
1963 (1992).

\bibitem{OND}
T. Ohta, H. Nozaki, and M. Doi, Phys. Lett. A {\bf 145} 304 (1990);
J. Chem. Phys. {\bf 93} 2664 (1990).

\bibitem {Ro}
D.H. Rothman, Europhys. Lett. {\bf 14} 337 (1991).

\bibitem{PT}
P. Padilla and S. Toxvaerd, J. Chem. Phys. {\bf 106} 2342 (1997).

\bibitem{PRL}
F. Corberi, G. Gonnella, and A. Lamura,  Phys. Rev. Lett. {\bf 81}, 
3852 (1998).

\bibitem{B94}
See, e.g., A.J. Bray, Adv. in Phys. {\bf 43} 357 (1994). 

\bibitem{Dino}
This solution is obtained by applying the method of characteristics
(See, e.g., R. Courant and D. Hilbert, 
``Methods of mathematical Physics'', J. Wiley  Interscience Publ. 1966).
Further details will appear elsewhere. 

\bibitem{Onu}
A. Onuki,  Phys. Rev. A {\bf 35} 5149 (1987).


\bibitem{mat}
K.Matsuzaka, T.Koga and T.Hashimoto, Phys. Rev. Lett. {\bf 80} 5441 (1998).


\bibitem{Bey}
C.K. Chan, F. Perrot, and D. Beysens, Phys. Rev. A {\bf 43} 
1826 (1991).

\bibitem{LLG}
J. L\"{a}uger, C. Laubner, and W. Gronski, Phys. Rev. Lett. {\bf 75} 
3576 (1995).

\bibitem{zan}
A.Coniglio, P.Ruggiero, and M.Zannetti, Phys. Rev. E {\bf 50} 1046 (1994).

\bibitem{RG}
A.J. Bray, Phys. Rev. B  {\bf 41} 6724 (1990).

\bibitem{nota2d}
For simplicity a two-dimensional system is considered, the extension to
arbitrary dimension being trivial. The calculated exponents ($\alpha_x, 
\alpha_y,\beta)$ do not depend on $d$.

\bibitem{Nota1}
The elimination 
of hard modes does not introduce any singular behavior 
in the scaling properties of the system \cite{RG}.

\bibitem{Sor}
See, e.g.,  D. Sornette, Phys. Rep. {\bf 297}, 
239 (1998).

\bibitem{July}
A. Wagner and J.M. Yeomans, cond-mat 9904033. 

\end{references}
\end{document}